\documentclass[journal]{vgtc}                

\ifpdf
  \pdfoutput=1\relax                   
  \usepackage{graphicx}                
  \DeclareGraphicsExtensions{.pdf,.png,.jpg,.jpeg} 
\else
  \usepackage{graphicx}                
  \DeclareGraphicsExtensions{.eps}     
\fi%

\graphicspath{{figures/}{pictures/}{images/}{./}} 

\usepackage{microtype}                 
\PassOptionsToPackage{warn}{textcomp}  
\usepackage{textcomp}                  
\usepackage{mathptmx}                  
\usepackage{times}                     
\usepackage{cite}                      
\usepackage{tabu}                      
\usepackage{booktabs}                  

\onlineid{1253}

\vgtccategory{Research}
\vgtcpapertype{application/design study}

\title{Visualization of Technical and Tactical Characteristics in Fencing}

\author{Mingdong Zhang, Li Chen, Xiaoru Yuan, Renpei Huang, Shuang Liu, and Junhai Yong}
\authorfooter{
	\item
	Mingdong Zhang, Li Chen, Renpei Huang, Shuang Liu, and Junhai Yong is with Tsinghua University.
	\item
	Xiaoru Yuan is with Perking University.
}

\shortauthortitle{Zhang \MakeLowercase{\textit{et al.}}: Visualization of Technical and Tactical Characteristics in Fencing}

\abstract{
 Fencing is a sport that relies heavily on the use of tactics. However, most existing methods for analyzing fencing data are based on statistical models in which hidden patterns are difficult to discover. 
 Unlike sequential games, such as tennis and table tennis, fencing is a type of simultaneous game.
 Thus, the existing methods on the sports visualization do not operate well for fencing matches. 
 In this study, we cooperated with experts to analyze the technical and tactical characteristics of fencing competitions. 
 To meet the requirements of the fencing experts, we designed and implemented FencingVis, an interactive visualization system for fencing competition data.
 The action sequences in the bout are first visualized by modified bar charts to reveal the actions of footworks and bladeworks of both fencers. 
 Then an interactive technique is provided for exploring the patterns of behavior of fencers. 
 The different combinations of tactical behavior patterns are further mapped to the graph model and visualized by a tactical flow graph. 
 This graph can reveal the different strategies adopted by both fencers and their mutual influence in one bout.  
 We also provided a number of well-coordinated views to supplement the tactical flow graph and display the information of the fencing competition from different perspectives. The well-coordinated views are meant to organically integrate with the tactical flow graph through consistent visual style and view coordination. We demonstrated the usability and effectiveness of the proposed system with three case studies. On the basis of expert feedback, FencingVis can help analysts find not only the tactical patterns hidden in fencing bouts, but also the technical and tactical characteristics of the contestant.%
} 


\CCScatlist{ 
 \CCScat{K.6.1}{Management of Computing and Information Systems}%
{Project and People Management}{Life Cycle};
 \CCScat{K.7.m}{The Computing Profession}{Miscellaneous}{Ethics}
}

\teaser{
	\centering
	\includegraphics[width=\linewidth]{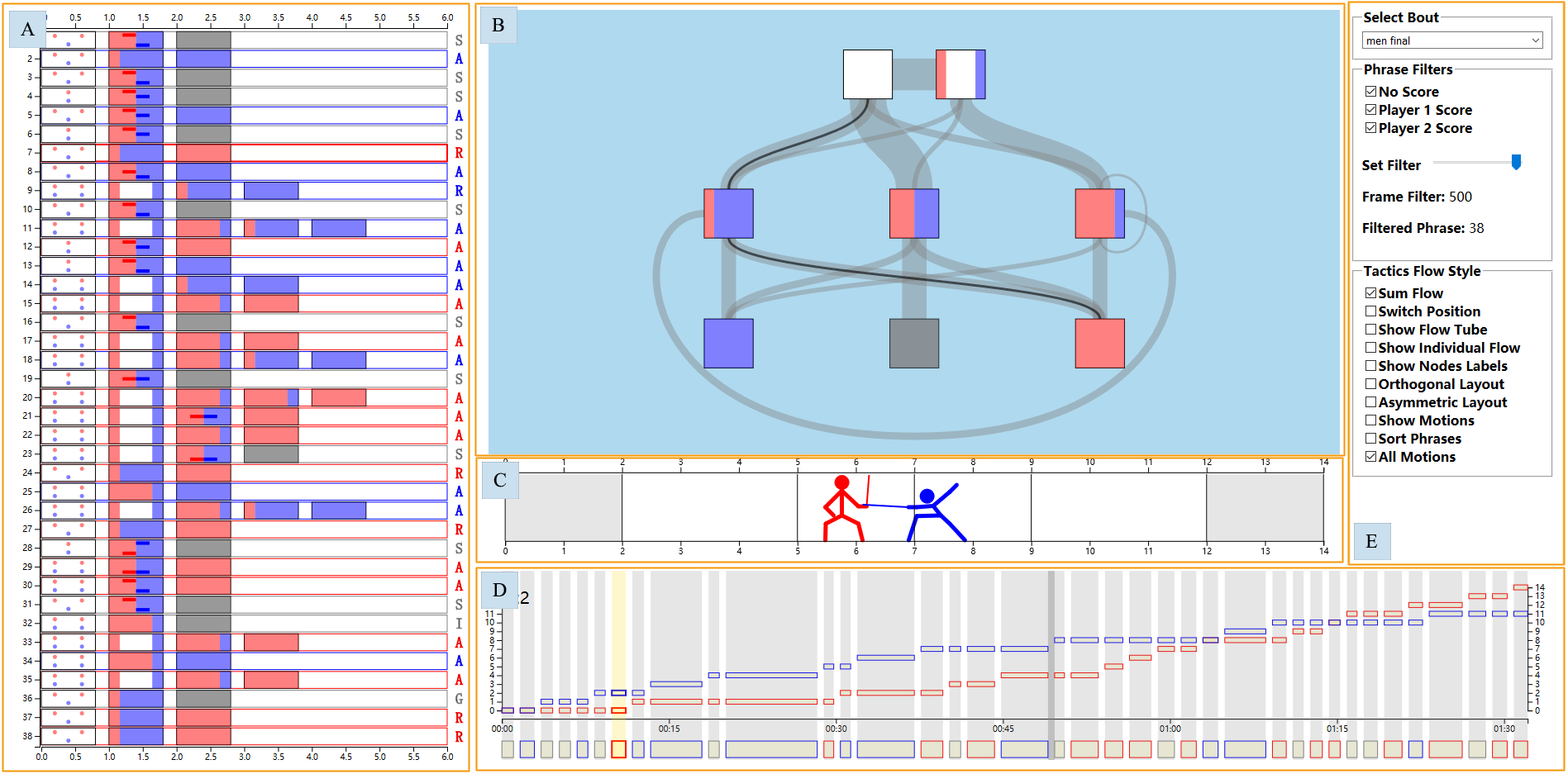}
	\caption{Men's Sabre Individual Golden Match of 2017 World Fencing Championship in FencingVis.
		 (A) Phrase List View shows the tactical usage of the fencers in each phrase. 
		 (B) Tactic Flow Graph View shows the statistics of the tactic usage in the whole bout. 
		 (C) Piste View shows the animation of the selected phrase. 
		 (D) Bout View shows the changing of the scores and the duration of each phrase. 
		 (E) Control Pannel provides set of controls to support interactive exploration.
	 }
	\label{fig:main}
}

\vgtcinsertpkg

\begin{document}


\firstsection{Introduction}

\maketitle

Fencing has a long history, and it is even one of the five activities featured in modern Olympics. However, participation in fencing has been relatively low. This phenomenon can be partly explained by the relatively high requirements for equipment and space, and more importantly, the difficulty of learning the game. In particular, fencing is difficult to learn mainly because the game is not easy to understand. Fencing is more abstract than other sports, such as table tennis and tennis. Let us take table tennis as an example. A table tennis match consists of several games. Each game is divided into multiple rounds, and each round is divided into multiple strokes. Different strokes can depict the different stroke techniques. By contrast, in a fencing bout, each phrase cannot be clearly described. In addition, fencing is a sport that relies heavily on the use of tactics, and the adoption of reasonable game strategies depending on the opponent. 

However, little research has focused on the data analysis of fencing. 
Some of the literature focused on statistical models\cite{tarrago2016complementary}, but these methods are unsuitable for discovering unknown tactic patterns. 
There also exists visual methods designed to help fully describe a fencing competition\cite{dentsu2016yuki}, but the technical and tactical characteristics of the game need to be clearly understood to assist in the training and the tactical arrangement of the competition. 

Many visualization methods are used to analyze sports competition data, but these approaches are largely unsuitable for fencing data. Most of the methods for sports data visualization are targeted toward multiplayer sports, such as soccer\cite{perin2013soccerstories,sacha2017dynamic,stein2018bring} and basketball\cite{goldsberry2012courtvision,chen2016gameflow}, in which data show completely different characteristics with those in fencing. 
Meanwhile, for table tennis, tennis, and most other single-player sports, the players' alternating actions generate data with hierarchical structure, which cannot be easily derived in fencing game.
Thus, the data visualization methods for single-player sports cannot be used for fencing data. The competition data generated for fencing competitions represent two related time-series data, and thus, visual methods to compare different sets of time series, such as those designed for bicycle sports, are suitable\cite{beck2016matrix}. Despite the similarities, the timing sequences of fencing data are not as simple as those in bicycle races. The time-series data of fencing competitions contain features of tactics, and these features cannot be extracted explicitly as those in table tennis and tennis, and they cannot be extracted automatically from time-series data, such as those in bicycle races. Due to the nature of fencing, none of the above methods can be used directly for fencing data analysis.

To fill the research gap, we cooperate with experts to analyze the technical and tactical characteristics that are regarded as either clear or fuzzy in fencing competitions, and subsequently summarize the requirements to explore the fuzzy problems through visual analysis. 
To meet these requirements, we design and implement an interactive visualization system for fencing competition data, which we call FencingVis. We first analyze the action sequences of fencers in a bout. Then, we extract the different sets of tactical behavior and construct a corresponding graph model to express tactical combinations. We design a tactical flow graph to show the constructed model and provide multiple interactive ways to explore it. We also provide a number of well-coordinated views to supplement the tactical flow graph. The viewing mechanisms can display the generated information of the fencing competition from different perspectives and organically integrate with the tactical flow graph through consistent visual style and view coordination. 

We demonstrate the usability and effectiveness of the proposed system with three case studies. According to experts, FencingVis can help them find not only the tactical patterns hidden in the fencing bout, but also the technical and tactical characteristics of the fencers. Apart from analyzing a professional competition, FencingVis can be used to teach fencing beginners and demonstrate tactics to fencing enthusiasts. 

The main contributions of the present study are as follows:

\begin{enumerate}
	\item 
	In-depth understanding of fencing data and requirement analysis, and from these aspects, a model for two-level data to represent the tactical and technical information of fencing; 
	\item 
	A novel visual design representing information at both tactical and technical levels, as well as multiple interactions and view associations to explore embedded patterns; and
	\item 
	Case studies using the data of a real competition to confirm the usefulness of the results of FencingVis to analysts.
\end{enumerate}

\section{Related Work}
Our work is mainly related to the analysis of fencing data and the visualization and visual analysis of sports data. Thus, we initially introduce the related work in these two areas.

\subsection{Analysis and Visualization for Fencing}
The existing analytical methods for fencing data are mainly at the technical level, and these efforts often analyze athletes from a biomechanical point of view. For example, by comparing the differences between excellent fencers and beginners, we can determine the factors with the greatest influence to provide guidance for the training of fencers\cite{chen2017biomechanics}. 
However, these methods are applied only at the technical level, which means that the use of tactics are generally not considered. Previous studies have used statistical methods for the time series analysis of fencing competition data\cite{tarrago2016complementary}, but the existing empirical model for data collection and game process analysis are summarized as a combination of several known patterns\cite{tarrago2015analisis}. 
We use this description of the bout for reference but selected to record the most primitive data, such as feet and blade movements. This type of recording level can reduce cost and information loss caused by introducing domain knowledge into the data acquisition process. The data abstraction work in the subsequent analysis process is considered to generate various benefits. First, the different disciplines of fencing are shown to have different behavior patterns, but all the actions can be recorded as the movement of blades and feet at the most basic level. As such, we can use a unified format to record data and apply logic in subsequent processing. In addition, if empirical models may change in the future, then we can modify the system logic without having to recapture data. 

Little literature has focused on the visualization of fencing data. Dentsu Lab Tokyo has conducted a fencing visualization project\cite{dentsu2016yuki}, but the visualization relies on a large number of sensors installed on fencers for data collection, which is not possible in real-life fencing matches. The main purpose of their visualization is to make fencing easy to understand and improve the aesthetics of the game. The former is one of our design goals, but our more important task is to provide the ability to explore fencing data interactively, not simply to show the collected data.

\subsection{Sports Visualization and Visual Analytics}
The visualization and visual analysis of sports data has developed vigorously in the past two decades, albeit with many challenges and opportunities. Basole et. al. attempted to summarize the two major difficulties of visualizing sports data\cite{basole2016sports}.
In addition to data complexity, the main issue of sports data visualization is the wide range of users whose information needs vary greatly. Previous work has often targeted the needs of a particular class of users, such as general sports enthusiasts, professional athletes and coaches, or psychological and physiological researchers. The design of FencingVis  is oriented toward professional and non-professional groups and aimed to meet their information needs at different levels. 

The visualization of sports data can be divided into two categories from the perspective of content analysis. The first category represents the full tournament or league season, in which data either show the points and rankings of each team during the season\cite{perin2016using} or provide support for game prediction\cite{vuillemot2016sports}. 
The second category is meant to analyze a single game, in which the situational dynamics of the game and the game information of two competing teams are presented. Some of the work is aimed at multi-player games, such as soccer and basketball, that focus on the spatial position of athletes and mine tactical information by analyzing spatial distributions and athlete tracks\cite{sacha2014feature,perin2013soccerstories}, whereas others focus on showing and analyzing the use of tactics or the characteristic abilities of individual athletes\cite{polk2014tennivis,wu2018ittvis}.

The present work falls into the category of single-player analysis, in which the two above mentioned working orientations are similarly covered in our scenario. TenniVis\cite{polk2014tennivis} uses scores and offers data to analyze amateur tennis matches. iTTVis\cite{wu2018ittvis} uses higher-level specialized data, such as placement and stroke techniques, to professionally analyze table tennis data. However, the above methods for content analysis are not applicable to fencing data because fencing has characteristics that differ from tennis and table tennis. First, in tennis or table tennis, every round ends with one player scoring. However, the case is different in fencing because some phrases (like round in tennis) can end with none of the fencers scoring (in sabre or foil) while some other phrases can end with both of the fencers scoring (in epee). Thus, fencing requires a different visual design. Second, the priority rules of fencing competitions are regarded extremely important and requires professional knowledge for judging. Non-enthusiasts may not understand why one fencer has scored with knowing the current priority. Therefore, the demonstration of priority is highly important in the visualization of fencing competitions. Finally, unlike sequential games, such as tennis and table tennis, fencing is a type of a simultaneous game with a different competition structure. 

\subsection{Other Relevant Visualization and Visual Analysis Methods}
The data we analyzed are fencing competition data. 
Besides the unique characteristics of fencing, these data also have some more generalized data characteristics. 
Therefore, we have also referred to some relevant visualization methods in our design process.
We designed the phrase list view to show the details of each phrase in the form of a list.
For the purpose of data analysis, users can choose different sorting methods. 
And when the sorting method changes, a smooth animation transition can help users to better maintain their mental map.
In order to achieve this effect, we draw lessons from the work of Gratzl et. al.\cite{gratzl2013lineup}. 
But their work is mainly to show the ranking, and we have more details, so we have added more elements to our list design, which needs to adjust the layout and animation design. 
In the design of tactical flow graph view, we also draw lessons from passing networks and transition network\cite{gudmundsson2017spatio}, but our nodes have different meanings, and more detailed designs have been added to the nodes and edges.

Our design is aimed to better explore and analyze fencing data, but the domain experts may not familiar with visualization and visual analysis. 
To facilitate the user, our design also includes many considerations of storytelling.\cite{figueiras2014narrative} 
For example, we use animation playback in piste view to connect the user with the scene quickly.
And in the design of view-coordination, we provide navigation of time view to connect the user with time and tactics.\cite{figueiras2014narrative} 

\section{Background and System Overview}
In this section, we present an overview of fencing, including the required data and the target of analysis. We also briefly describe the main components of the system.

\subsection{Background}
Fencing is one of the representative activities of the Olympic Games, and it evolved from the swordsmanship techniques used for the military combats and duels of the Cold War era. Fencing comprises the three disciplines of epee, foil, and sabre, in which scores are earned by hitting the active body parts of an opponent. The basic techniques of fencing are divided into offensive and defensive techniques. Offensive techniques include attack, riposte, feint, lunge, beat attack, disengage, compound attack, continuation/renewal of attack, and flick. Defensive techniques include parry, circle parry, counter attack, and point-in-line. These techniques are learned through limited combinations of blade and feet movements. There are two types of fencing competitions: individual and group. In an individual match, the fencer who first scores 15 points wins the game. After a fencer scores 8 points in a sabre match, the two sides take a minute off before the game is continued. 

A fencing match is called a \textbf{bout}, which consists of several \textbf{phrases} (i.e., a set of related actions and reactions) . At the beginning of each phrase, two fencers stand behind the two on-guard lines at both sides of the \textbf{piste} (game field of fencing) and perform their actions after the referee gives the signal. Each phrase can be ended with one fencer scoring, both fencers scoring (epee), or neither of the two fencers scoring (sabre or foil).

Unlike other sports, fencing has a special priority rule\cite{roi2008science}. 
This rule is applied to sabre and foil.
The fencer who initiated the attack first gains the priority, and each attack will lead to the exchange of the priority. 
When two fencers hit each other at the same time, the one with the priority scored. 
If it is not possible to accurately judge the priority, both fencer will not score point in this clash.
Judging the priority is not trivial, so showing the current priority is also one of the important considerations in our system design.

\subsection{Data Description}
Owing to the fast-pace characteristic of fencing, the detailed real-time recording of the match is difficult to conduct. Moreover, to avoid interfering with competitors, it is not convenient to install sensor devices. The existing method of analyzing fencing competitions is achieved by taking videos of a match from which sports data are extracted. In general, the accuracy of a game video is 30 frames per second (fps), which means that data are video-recorded frame by frame at a time accuracy of 1/30 seconds. For each data frame, the listing attributes are recorded (see \autoref{tab:data}).

In the process of data-marking the game videos, continuous footwork does not necessarily mean effective segmentation. Thus, after consulting domain experts, we use the start time of the next action as the segmentation point of two continuous actions. Specifically, for the continuous forward movement, we use each front foot off the ground as the segmentation point. For the continuous backward movement, we use each the rear foot off the ground as the segmentation point.

\subsection{Requirement Analysis}
On the basis of extensive discussions with field experts, the characteristics of fencing that need to be considered in the visual design are as follows:

\begin{itemize}
	\item Fencing is not as easy to understand as tennis and table tennis and similar games, and non-enthusiasts often find it difficult to readily understand fencing bouts. The visual design should therefore be able to contribute to the enhanced understanding of specific data users.
	
	\item The current information generated in most sports competitions are generally clear, but the case of fencing is different because the most important information, such as priority, should be considered. Viewers and data users with different experiences may have different understanding of gaming decisions, and this scenario implies that a common visualization is seldom achieved. 
	
	\item Most sports are bound to end each round with one side scoring, but this is not the case in fencing. Fencers may either score both or neither in a phrase, and this scenario needs to be reflected in the system designs. 
	
	\item The use of tactics is more important in fencing compared with other sports that place more emphasis on adaptability. Furthermore, fencing tactics are often planned in advance before each phrase, and thus, it is more valuable to show the impact of this strategy on the bout. 
\end{itemize}

\begin{table}[]
	\centering
	\caption{Data Description}
	\label{tab:data}
	\begin{tabular}{|p{1.5cm}|p{6.5cm}|}
		\cline{1-2}
		Bout ID  &  The ID of the bout to which this event belongs.\\	\cline{1-2}
		Phrase ID  & The ID of the phrase to which this event belongs.\\	\cline{1-2}
		Frame  &  The frame at which this event occurs.\\	\cline{1-2}
		Footwork1  & Beginning or finishing of forward, backward, or lunge of fencer 1\\ 	\cline{1-2}
		Footwork2  & Beginning or finishing of forward, backward, or lunge of fencer 2\\ 	\cline{1-2}
		Bladework1  & Beginning or finishing of attack, parry, riposte, or counter attack of fencer 1\\ 	\cline{1-2}
		Bladework2  & Beginning or finishing of attack, parry, riposte, or counter attack of fencer 2\\ 	\cline{1-2}
		Attack1  & Attacked position of fencer 1\\ 	\cline{1-2}
		Attack2  & Attacked position of fencer 2\\ 	\cline{1-2}
		Parry1  & Parried position of fencer 1\\ 	\cline{1-2}
		Parry2  & Parried position of fencer 2\\ 	\cline{1-2}
		Confrontation& Confrontation position of the two fencers on the strip.\\ 	\cline{1-2}
		Result  & Result of this phrase, which is given by the referee.\\ 	\cline{1-2}
		Score  & Record which fencer scored or none.\\ 	\cline{1-2}
	\end{tabular}
\end{table}

Based on these discussions, we summarized the requirements of our applications as follows:
\begin{itemize}
	
	\item (R1) Show how the bout changes over time.
	\begin{itemize}
		\item (R1a) Show changes in scores.
		\item (R1b) Show the length of each phrase.
		\item (R1c) Show the changing of priority ownership. 
	\end{itemize}
	\item (R2) Show a detailed comparison of phrases at both tactic and technical levels. 
	\begin{itemize}
		\item  (R2a) Show the applied tactics of both fencers in different phrases.
		\item (R2b) Show the technical details of the selected tactics.
	\end{itemize}
	\item (R3) Show how the tactics of both fencers are used in the entire bout. 
	\begin{itemize}
		\item  (R3a) Provide a summarized view of the tactics use during the bout.
		\item  (R3b) Map the summarized view with the listed details of each phrase.
	\end{itemize}
	\item (R4) Conduct exploratory pattern discovery and result communication. 
	\begin{itemize}
		\item  (R4a) Arrange information according to user interaction to aid in pattern discovery.
		\item  (R4b) Represent clearly the discovered pattern to aid in the communication of users.
	\end{itemize}
\end{itemize}

\begin{figure*}[tb]
	\centering
	\includegraphics[width=\linewidth]{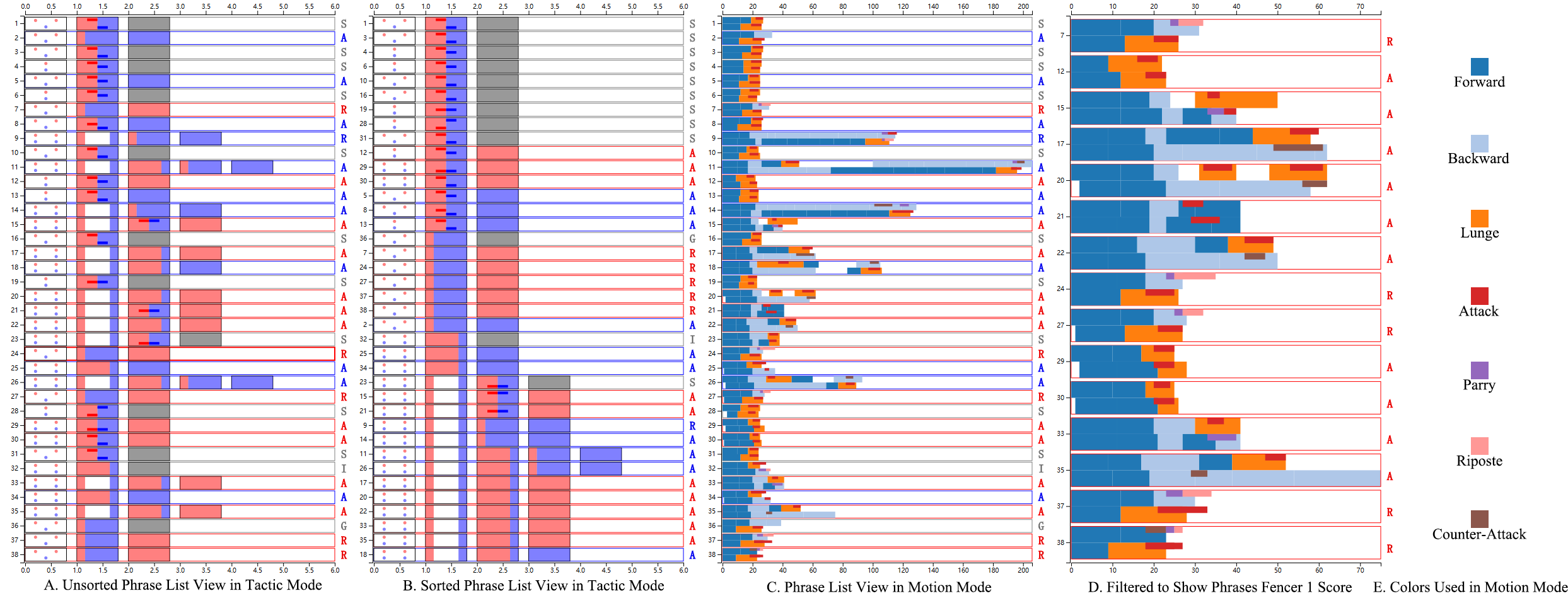}
	\caption{The phrase list view presents the details of each phrase, and the two modes of this view separately represent two different abstract levels.}
	\label{fig:listview}
\end{figure*}

\subsection{System Overview}
Our system consists of four views and a control panel to support analyzing the data on a match from
four different perspectives, as shown in \autoref{fig:main}.
\textbf{Bout View} shows how the bout changes over time (\textbf{R1}).
\textbf{Phrase List View} shows the details of every phrase form both tactical and technical level in the form of a list (\textbf{R2}).
\textbf{Tactical Flow Graph View} shows the overall statistics of the use of tactics of the fencers (\textbf{R3}).
\textbf{Piste View} show the details of selected phrase in the form of an animation on the piste.
\textbf{Control Panel} provides set of controls to support the interactive exploration (\textbf{R4}).

\section{Visualization Design}
We use consistent visual styles in the whole application:
\begin{itemize}
	\item \textbf{Color:} We use red and blue as the representative colors of fencer 1 (left side) and fencer 2 (right side). 
	This principle is embodied in all the views.
	Furthermore, in the glyphs expressing the tactical combination of two fencers, the proportion of their respective colors reflect who having the priority (\textbf{R1c}).
	\item \textbf{Layout:} For all visual elements used to compare two fencers, we try to arrange them horizontally, and keep the information related to fencer 1 on the left side and information related to fencer 2 on the right side, which is consistent with the actual positions of the fencers. If it really needs to be arranged up and down, information related to fencer 1 is always on the top.	
\end{itemize}

\subsection{Bout View}
Most game data naturally have time attributes, and both tactical and technical analysis are needed to consider the impact of time. The influence of time is reflected in two aspects. 
First, the different stages of the game and the psychological and physical changes of athletes significantly affect the game results. 
Second, the use of tactics has time-dependent characteristics. A fencer executes tactics based on the previous ones that he or she and his or her opponent have used over a certain period. The choice of tactic repetition or conversion therefore needs to be determined on the basis of the characteristics of the previous phrase and the tactics used by the opponent. 
Finally, analysts need to know how the competition has changed over time. The bout view (\autoref{fig:main}D) is therefore designed to show this information.

The bout view mainly shows three elements: time, score change, and phrase duration. 
We use a tailored step-chart to show the variations in scoring according to time (\textbf{R1a}). 
In the chart, the x-axis mapping represents game time whereas the y-axis mapping represents scores.
The red and blue rectangles represent the scores of both fencers, and the two scores naturally overlap (purple rectangle) when equal. 
To visually compare the duration of each turn (\textbf{R1b}), we add a horizontal rectangle below the x-axis to show each phrase. The color of the rectangle indicates the fencer who scores in the phrase, and a gray rectangle indicates that neither fencer has scored. The upper and lower views are designed to correspond to each other to help users visually observe the relationship of the three attributes. Considering that the break called in between the first and second half of the match often has a great impact on the course of the game, we use a vertical line to emphasize this split moment. To depict the selection of a phrases, we use a gray background to reflect the selected phrases.

\textbf{Description:} The game time in our viewing scheme does not exactly correspond  to actual time. Considering that the time required by fencing phrases accounts for a small proportion of actual time, the view becomes very sparse if mapped directly. As such, we map the time of the phrases directly, and the combined time of two adjacent rounds represent an interval of 1 second.

\subsection{Phrase List View}

\begin{figure*}[tb]
	\centering
	\includegraphics[width=\linewidth]{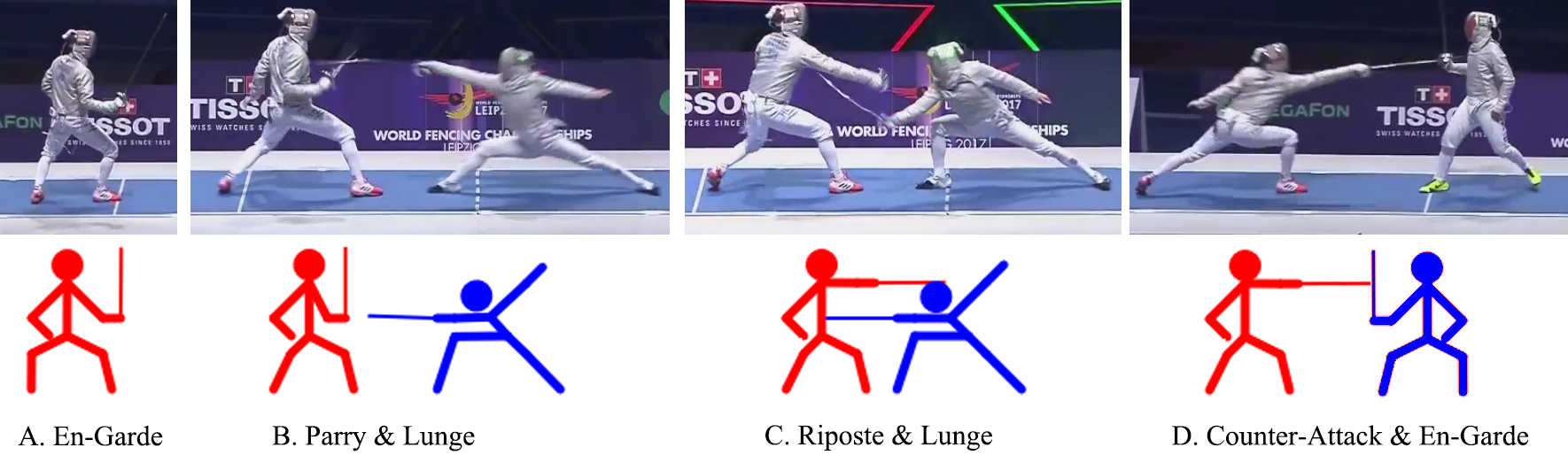}
	\caption{The pose of the fencers on the piste is abstracted and designated as four glyphs}
	\label{fig:glyph}
\end{figure*}

The phrase list view (\autoref{fig:main}A) presents the details of each phrase (\textbf{R2}), and the two modes of this viewing scheme separately represent two different abstract levels. 

In the \textbf{motion mode}, the motions of two fencers in each phrase are displayed as a function of time from the beginning of the phrase (\textbf{R2b}). Each listed item is subdivided into two lines (up and down) to describe the actions of the two fencers, with the upper and lower parts for the left and the right fencers, respectively. Bar charts are used to describe the motions of fencers in each phrase, in which the horizontal axis corresponds to time duration, as described by the number of frames. We use the data from 30 fps video, and thus, the scale is the same.The corresponding actual timespan can be viewed in the bout view when this is needed. The tall bars in the bar charts represent feet movements, whereas the short bars embedded in the tall bars represent bladeworks.
Various colors are assigned for different type of actions, as illustrated in \autoref{fig:listview}E. 

In the \textbf{tactic mode}, the abstracted tactic nodes are displayed (\textbf{R2a}). Specifically, the motions of two fencers in a phrase is abstracted as a sequence of tactic nodes. The motions can be described as follows: 

\begin{enumerate}
	\item Start (\textit{S}): The state when the referee issues a start command, and both fencers behind the en-garde lines (or the position where the last phrase interrupted). The start state is the first state of each tactic sequences, and the source node of the tactical flow graph described in the next section. 
	
	\item Forward-Forward (\textit{FF}): The state when both fencers attack forward at the same time. The game enters the \textit{FF} state in either of the two scenarios: the standard \textit{FF} when both fencers initiate forward movements simultaneously or the scenario in which both fencers make backward movements and switch to \textit{FF} states simultaneously. 
	
	\item Backward-Backward (\textit{BB}): The state of simultaneous retreat. Considering that both fencers will likely step forward at the start of a fencing phrase, our decision to include the \textit{BB} state is not literally based on actual backward movements, but whether the \textit{BB} state has been planned before the phrase. The \textit{BB} state usually occurs when both fencers move a step or two forward, then switch to \textit{BB} to back away. When a fencer pauses as he or she moves forward and subsequently decides to further move forward, we also designate this phenomenon as a \textit{BB} state, i.e., the movement was a fast-changing forward.
	
	\item Backward-Forward (\textit{BF}): The state when the left fencer chooses a backward movement, whereas the right fencer chooses the forward movement.
	
	\item Forward-Backward (\textit{FB}): The state when the left fencer chooses a forward movement, whereas the right fencer chooses a backward movement.
	
	\item Fencer 1 score (\textit{1}): The state when the left fencer scores. 
	
	\item Fencer 2 score (\textit{2}): The state when the right fencer scores.
	
	\item Simultaneous (\textit{=}): The state when both fencers hit each other simultaneously but no scores are given.
\end{enumerate}

The glyphs of the tactic nodes are designed to allow users to easily understand their meanings. 
The white rectangle represents the \textit{S} node, whereas the red, blue, and gray rectangles separately represent nodes \textit{1}, \textit{2}, and =, respectively. 
The color scheme is consistent with our basic design principle wherein red and blue represent that dominance of the left and the right fencers, respectively. 
In accordance with this deign principle, the glyphs of the four tactic nodes are designed with red and blue to denote two parts in different areas. However, the tactic nodes only show tactical information. Thus, the nodes are designed to represent additional details. 
In the \textit{S} node, one dot or two dots are separately used to show one step or two steps forward performed by both fencers. 
In the \textit{FF} nodes, two small rectangles depict the attack positions of both fencers (as shown in \autoref{fig:listview}A and B). 

In both  modes, the labels to the left of each phrase box represent the index of the current phrase, which can be quickly retrieved. 
Meanwhile, the labels on the right show the outcome of each phrase. 
We use the first letter of the word to depict the call of the referee (i.e., A for attack, R for riposte, and S for simultaneous). 
The phrases are arranged as a series of rows from top to bottom in the order of the game, in which each row describes the information from one phrase. 
To support the interactive exploration of data, we provide different sorting approaches, as shown in \autoref{fig:listview}B. By sorting the phrases according to different rules, such as the order in which they occurred or different priorities of the tactical combinations, users can easily find the different features that represent the same tactic nodes and sequences (\textbf{R4a}). 

To highlight the score of a phrase, we use the color of the two fencers (red and blue) to render the border and text of the phrases. 
The non-coloring of fills is intended to avoid interference with the display information of internal details, i.e., if the internal details are set to translucent, then the color inconsistencies can lead to confusion. 

The items shown in motion view are affected by the filter setting, including those for the score and duration filters in the control panel as shown in \autoref{fig:listview}D.

\begin{figure*}[tb]
	\centering
	\includegraphics[width=\linewidth]{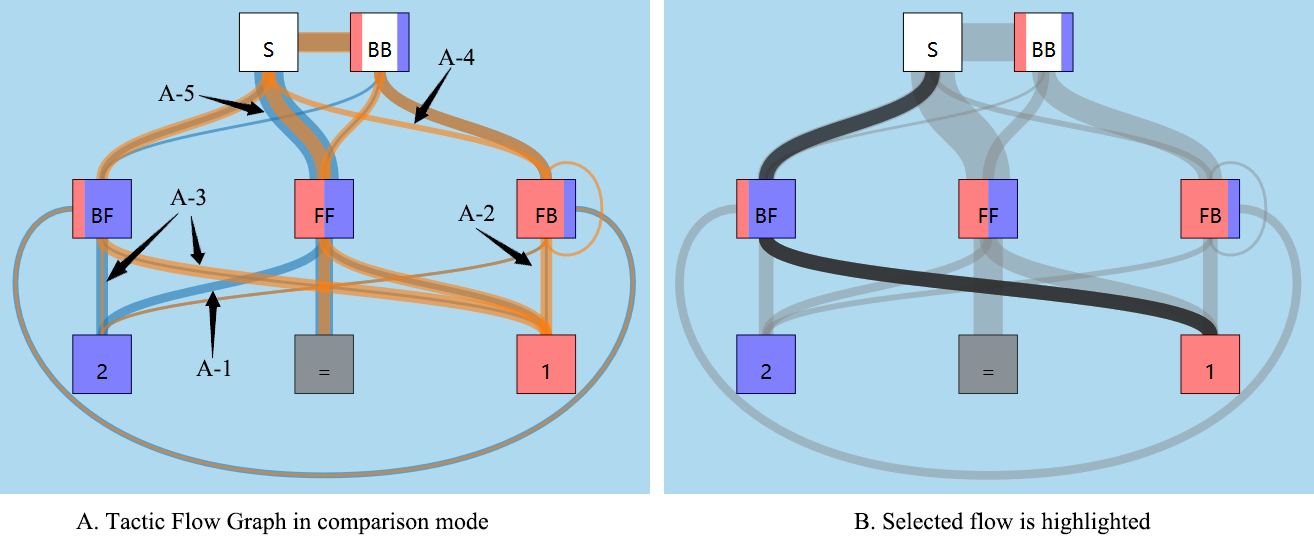}
	\caption{Tactical flow graph view of Men’s Sabre Individual Golden Match of 2017 World Fencing Championship.
		Each glyph in the graph indicates a state in the competition. 
		The colors in the glyphs indicates the advantages and disadvantages of both fencers in this state. 
		The greater the color ratio of a fencer, the greater the advantage of the fencer.
		In the first row, the white 'S' indicates the beginning state, at which the situation between the two sides is equal. 
		Glyph of 'BB' state is white in the middle and has a narrow red and blue color on both sides, indicating that both fencers retreat at the same time. 
		In the second row, the red-blue equivalent 'FF' indicates that both fencers are advancing at the same time. 
		Correspondingly, 'BF' and 'FB' indicate that one fencer is advancing and the other is retreating respectively. 
		The three glyphs in the third row indicate blue side scored, on one scored and red side scored respectively.		
	}
	\label{fig:case1}
\end{figure*}
\subsection{Tactical Flow Graph View}

The bout view and the phrase list view are designed to allow users to understand the game more clearly with no much statistical summaries. Thus, to allow analysts to have a relatively deeper understanding, we designed a tactical flow graph to show summary of the tactics used by the fencers in the bout (\textbf{R3a}). 

In analyzing the tactical information of a fencing competition, we initially converted the collected time series to a tactical graph model. We consulted professional fencing coaches and athletes to devise a series of conversion rules. The modeling process needs to introduce some domain knowledge that cannot be directly retrieved at the level of data conversion. For example, to check whether the fencer has chosen forward or backward tactics, we need to look at the behavior after the first two or more steps. The use of strategy itself is a process of deception and anti-deception\cite{roi2008science}. A fencer usually uses two-step lunges, but one-step lunges to bring about a sudden attack is also used. Apart from the one-step lunge, a backward movement can also be performed after a two-step forward.

The design of the tactical flow graph view (\autoref{fig:main}B) is based on the graph model we built upon the game data. After the in-depth discussions with experts, we arranged the 8 nodes according to the following criteria:

\begin{itemize}
	\item The designed view shows the progress of the game from top to bottom. The nodes are naturally arranged in three layers as follows: The first layer, which contains nodes \textit{S} and \textit{BB}, represents the start of each phrase. The second layer, which contains nodes \textit{FB}, \textit{FF}, and \textit{BF}, represents the middle stage of the phrase. The third layer, which contains nodes \textit{1}, = , and \textit{2}, represent the end of the phrase. All data entries in the graph flow only from the upper layer to the lower layer or between the same layer.
	\item The node layout of the viewing scheme in the horizontal direction denotes the advantages gained by the fencers. 
	The nodes are arranged in three columns as follows: The left column indicates that the right fencer dominates in terms of scoring or priority (the left fencer is retreating). The right column indicates that the left fencer dominates given the same conditions. The middle column represents the balance of power between the two fencers. 
	\item Although the design principle is implemented in its entirety, tradeoffs are considered to ensure that the view schemes are clean and tidy. For instance, nodes \textit{S} and \textit{BB}, which should have been arranged up and down if the above rules are followed strictly. However, this would bring in substantial overlaps. We therefore arranged the two nodes as left and right, but close to one another in the same region, and this design is consistent with the above rules at the regional level. This approach also makes the flow \textit{S-BB} to be in a focused position (center on upper side) to show their relative importance to analysts (i.e., see case study discussion).

\end{itemize}

\subsubsection{Orthogonal Layout}
The designing of our tactical flow graph fully considers the actual physical scene, so that experts can more intuitively understand the information it expresses.
But the experts also need to compare the tactical flow graph of the two halves or of different matches.
For the comparison between the first and second half, as there are just two items to be compared, we can directly use two translucent color to show them, as shown in \autoref{fig:case1}A.
But when more than two graphs are compared, this superposition can cause confusion.
We also tried to lay the flows side-by-side, but because of the intersection of current designs, this may also cause visual confusion.

To better illustrate the comparison of multiple tactical flow graphs, we provide an alternative orthogonal layout, as shown in \autoref{fig:comparison}.
We arrange all the nodes in an orthogonal grid, with the \textit{S} node at the center as the origin, and all the flow flows from the center to the periphery.
Because each node of orthogonal layout can only be adjacent to four nodes, it is not enough to show all the inflow and outflow, so as a cost, we introduce redundant nodes.
Since this layout is only intended to show the comparison of tactical flows and does not focus on the overall path, the introduction of redundant nodes does not impose a burden of understanding.
But we make sure that the flow direction is from inside to outside, and that all flows occur only once.

\subsubsection{Operations}
The above view shows the information at the highest level. 
Users can find some patterns with higher abstraction level from this view, and further exploration is often needed for the content of interest.
To do this, we designed a series of interactions to show more detailed information and related statistics.

When the mouse moves to a node on the tactical flow graph, the flow through that node is highlighted.
This is to help users better observe the relationship between each node and the flow.
Similarly, when the mouse moves over a segment of data flow, the associated data flow is highlighted to help the user quickly observe the source and flow of the segment, as shown in \autoref{fig:case1}B.

When the mouse moves to the \textit{FF} node, the upper left corner and the lower left corner of the view respectively displays the matrix of attack positions and the matrix of forward steps, which can help the user to get the patterns of the technical details of the fencers.

\subsection{Piste View}

Although the phrase list view shows the details of each phrase, it has a time-dimensional presentation with certain limitations, and one of these is the inability to reflect the position information on the piste, which is not intuitive for domain experts. 
To resolve this deficiency, we design the piste view to show the information of each phrase in the form of animation. 

Our animation design mainly reflects two aspects of information: the position on the field and the postures of both fencers. 
These two relevant aspects of information can be disassembled. 
We use a more flexible design to animate the two layers and subsequently determine the changes in the two types of information, which are stacked together to show the information of a phrase.
Animating the position is relatively simple to implement, i.e., the position of the glyph is driven by the position of recorded data. 
In animating the pose, we abstract the pose of a fencer on the piste and designate them as four glyphs on the basis of the observed games and the suggestion of domain experts. 
The four glyphs (shown in \autoref{fig:glyph}) are as follows:
\begin{itemize}
	\item \textbf{En-garde glyph} is used to represent the en-garde posture.
	\item \textbf{Lunge glyph} is used to represent the lunge posture.
	\item \textbf{Parry glyph} is used to represent the parry posture.
	\item \textbf{Riposte glyph} is used to represent both the riposte posture and the counter attack posture.
\end{itemize}
\begin{figure*}[tb]
	\centering
	\includegraphics[width=\linewidth]{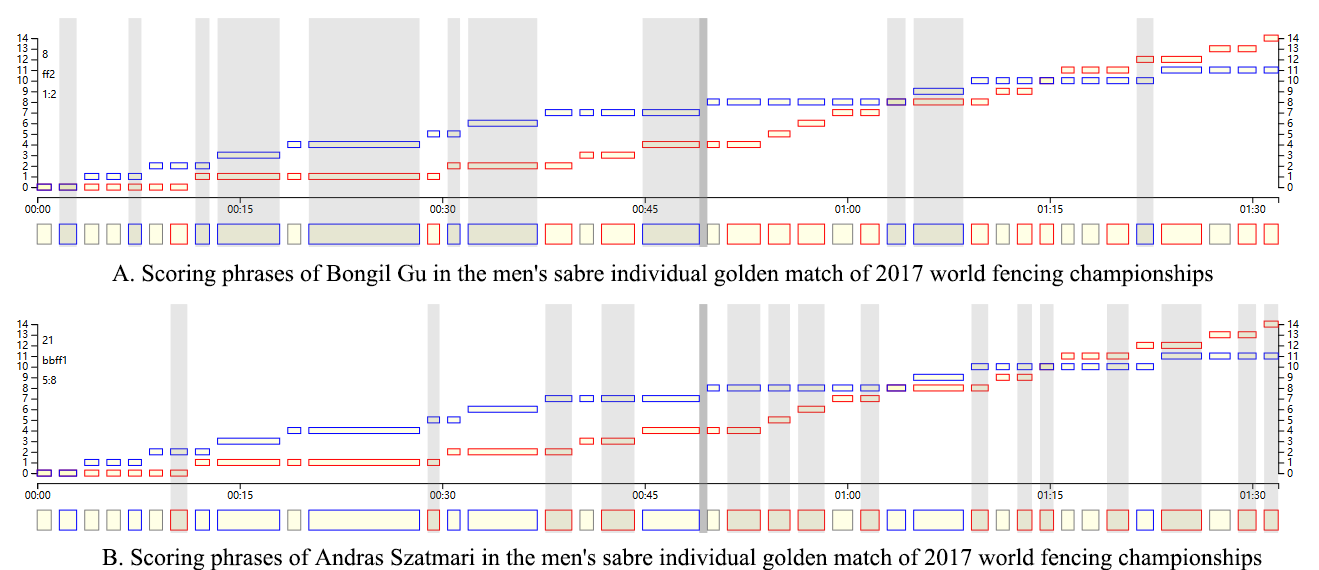}
	\caption{Bout View of Men's Sabre Individual Golden Match of 2017 World Fencing Championship with highlighted scoring phrase of each fencers.}
	\label{fig:boutview}
\end{figure*}

\subsection{Control Panel}
To meet the basic requirements of interactive data exploration, we provide some control components that are mainly used to filter and change the display mode of the viewed data, as shown in \autoref{fig:main}E. First, we provide a drop-down menu to select the game intended for analysis. 
We also provide filters for the phrases, which correspond to results and time dimensions. 
The resulting filters are used to select the combinations of the scoring phrase of the fencer 1, the scoring phrase of the fencer 2, and the no-score phrases. 
The time filter selects a time threshold through a time bar, and the phrases with duration turns that are longer than the thresholds are filtered out, thus leaving only short phrases. The effects of the two filters can be superimposed, and the results are updated synchronously on the bout view and the phrase list view. 

To support keeping the number of filtered items in mind, the filtering threshold and the number of filtered results are displayed simultaneously. The display mode control for the data flow can be selected by users in three modes: display the entire game, compare the first and second halves of the game, or exchange the position of fencers (i.e., a scenario for comparing the different games of the same fencer).

If a fencer is simply positioned at different sides of the game field, then a comparison of the data flow graph will not be as intuitive as expected; switching the view in the same direction is an easier option. The background of the data stream can also be selected in viewing or unviewed mode. For instance, users who simply want to experience using the system without having to analyze the game can select the background to quickly view the different data flow nodes.

\subsection{Cross-view Analysis}
The interactive exploration of data is mainly realized through the association of views. The main view associations include the following:

\begin{itemize}
	\item After the user modifies the filter settings, the phrase list view and the bout view are both updated synchronously. The former displays the filtered results only, whereas the latter highlights the filtered results with a gray background.
	\item When the mouse hover on the items in the phrase list view or in the bout view, the corresponding phrase in the other view highlights the border for selection. The data flow of this phrase is also displayed synchronously in the data flow graph view. 
	\item User can click the items in the action view or bout view to trigger the animation of the corresponding phrase displayed in the piste view. 
\end{itemize}

\section{Case Studies}
We demonstrate the usability and effectiveness of the proposed system with three case studies.
The three cases are based on the semi-finals and finals of the Men's Sabre Individual Golden Match of 2017 World Fencing Championship.
We explore from three perspectives of single match analysis, comparison of multiple matches and comparison of different matches of the same fencer.
In this process, domain experts participated in the whole process and put forward assumptions and guidance for our analysis in real time.

\subsection{Men's Sabre Individual Golden Match of 2017 World Fencing Championship}

\begin{figure*}[tb]
	\centering
	\includegraphics[width=\linewidth]{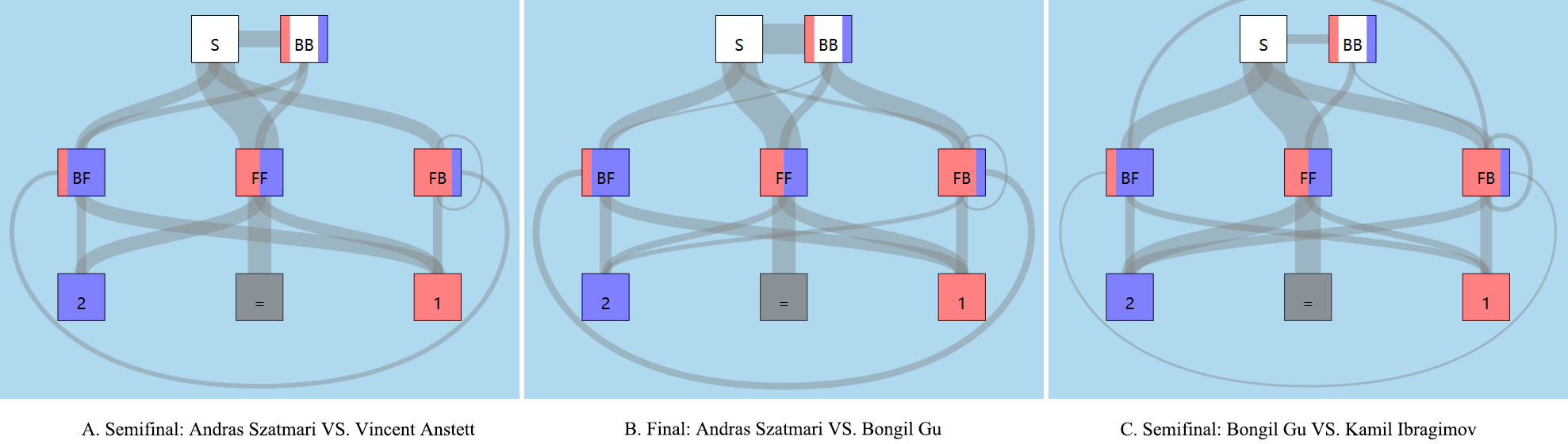}
	\caption{Comparison of tactical flow graphs in the semi-finals and final of fencing world championships in 2017}
	\label{fig:sample}
\end{figure*}

We analyze the final match of Szatmari and Gu in the Men's Sabre Individual Golden Match of the 2017 World Fencing Championship. A quick look of the bout view shows Gu leading in the first half, then Szatmari reversed the game play in the second half. The wins and losses of the game are related to changes in strategies in the first and second halves of the game. Users can switch the tactical flow graph to the half-court view, as shown in \autoref{fig:case1}A.

Most of Gu's scores in the first half were enabled by nodes \textit{FF} and \textit{BF} (shown by blue flow), but the scores were significantly reduced in the second half (shown by orange flow). At the same time, the sources of Szatmari's second-half main scoring are from nodes \textit{FB} and \textit{BF}. By summarizing the four obvious flow changes, we present the following preliminary conclusions:

\begin{enumerate}
\item Gu's scoring from both sides choosing direct attack were reduced in the second half  (\autoref{fig:case1}A-1). 
\item Szatmari increased his scores in the situation that he moving forward while his opponent moving backward in the second half (\autoref{fig:case1}A-2). 
\item Gu's forward movement with Szatmari's backward movement mainly contributed to Gu's scores in the first half, but the scenario shifted to Szatmari's favor in the second half (\autoref{fig:case1}A-3). 
\end{enumerate}

Based on the above observation results, together with domain experts, we are trying to find out the deeper reasons. Thus, we shift our attention from the lower half to the upper half of the tactical flow graph view. 
We can intuitively see that the \textit{S-FB} flow increased in the second half (\autoref{fig:case1}A-4) while the \textit{S-FF} flow decreased in the second half (\autoref{fig:case1}A-5). Together, they reflect Szatmari's forward tactic usage increased and Gu's backward tactic usage decreased in the second half of the bout.

Another obvious change is the \textit{S-BF} flow, which increased significantly in the second half. To further analyze the case of the \textit{BF} node, we switch back to the sum flow mode of the tactical flow graph view and select the \textit{S-BF} segment, as shown in \autoref{fig:case1}B. At this point, most of the \textit{S-BF} flows end at node \textit{1}, which correspond to Szatmari's scoring, and most of which occurred in the second half. We can make the following assumptions about the course of the game:
\begin{enumerate}
\item Gu led the game by earning points in the first half by relying on his strong offensive ability. Szatmari failed to handle Gu's attacks regardless whether the forward or backward tactic was applied.
\item After the game break, Szatmari adapted and started to retreat to counter Gu's attacks. Szatmari was successful and he scored many times, as depicted by the \textit{BF-1} segment. 
\item As a consequence of being countered many times, Gu began to hesitate on his attacks and opted for more backward movements. This scenario is reflected in the decrease in simultaneous-attacks and the increase in Szatmari's forward movement and Gu's retreat.
\item Collectively, the above factors led to Gu's defeat.
\end{enumerate}
To confirm the above assumptions, we locate the start phrase of the second half in the game view and check the details of a few succeeding phrases in the motion mode of the phrase list view (\autoref{fig:listview}C). In the phrase list view, we see that Szatmari consecutively earned offensive points in the phrases at the beginning of the second half. This finding differs with our previous assumptions. At the same time, Szatmari's backward scorings were earned at the end of the game. As such, we redefine our understanding of the game as follows:
\begin{enumerate}

\item Gu led the game by earning points in the first half by relying on his strong offensive ability. 
\item However, although Gu's attacks were sharp, his physical energy was greatly consumed in the first half. Thus, at the start of the second half, Gu's attacks began to falter, which opened opportunities for riposting from the opponent. 
\item Gu changed his strategy, and his retreats increases. However, Gu's ability to retreat was insufficient, and his scoring was eventually surpassed by his opponent.
\item Gu had no choice but continued attacking, but his opponent had detected the decline in his speed, and continuously retreated to counter Gu's attack. As a result, Szatmari finally won the game.
\end{enumerate}

We summarized this match. Gu's offensive ability was very strong, but this affected his physical bearing, and this condition led to his unsustainable attacks in the second half of the game. To sustain his gaming advantage, Gu should have focused on his physical strength to ensure that his offensive ability does not decline. Other gaming aspects, such focusing on the short board, might have also been an effective method for Gu to overcome his declining attack ability. Szatmari's ability can be regarded as relatively average, but his timely discovery of the changed state of his opponent in the second half proved to be a reasonable adjustment strategy. Szatmari eventually won the game. 

The scoring phrases of both fencers were also analyzed in the bout view. All of the long-duration phrases contributed to Gu's scoring (\autoref{fig:boutview}A), whereas Szatmari's scoring were all in the short-duration phrases (\autoref{fig:boutview}B). This finding also confirms that the technical ability of Gu is better than that of Szatmari, but the latter won the game because of his reasonable use of tactics.

\begin{figure*}[tb]
	\centering
	\includegraphics[width=\linewidth]{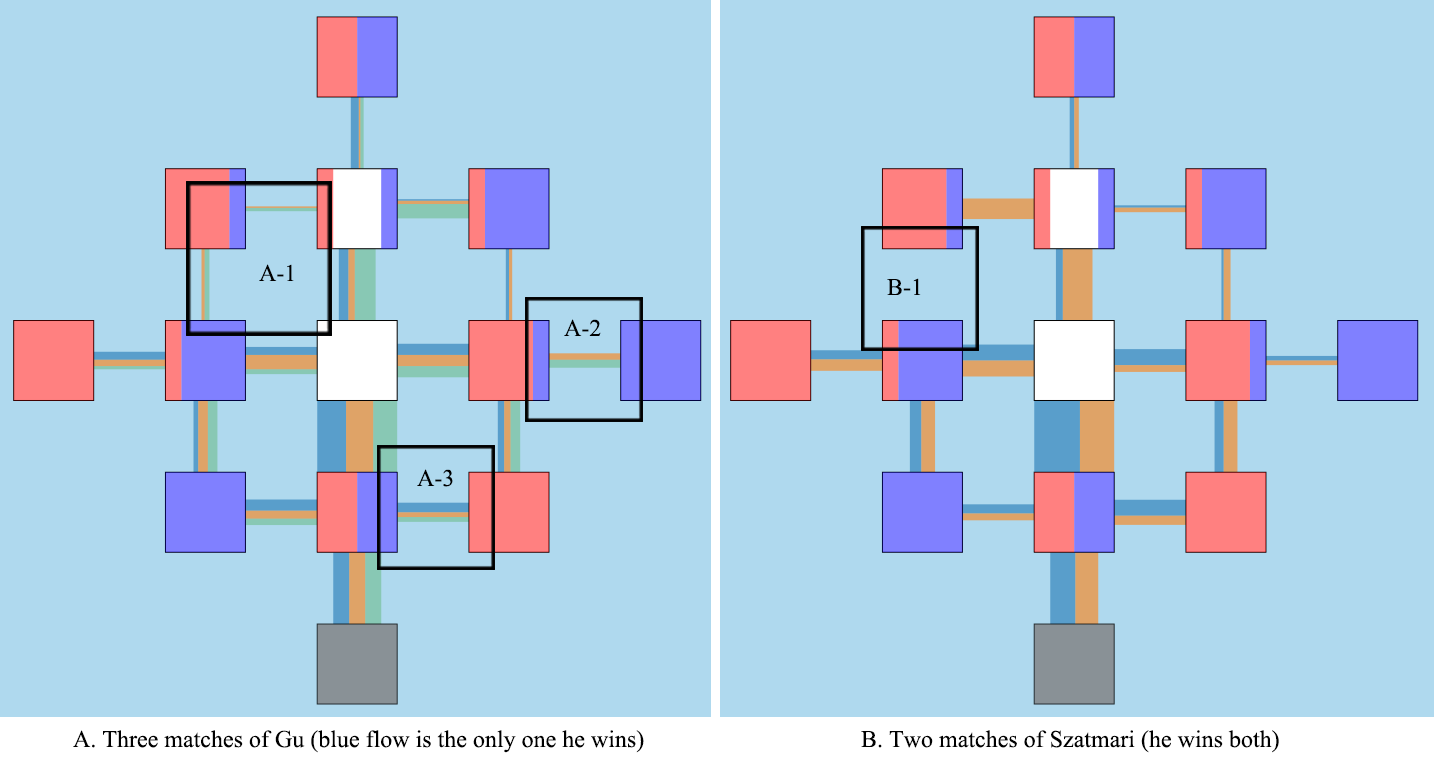}
	\caption{Our system also supports the analysis of the characteristics of a fencer by loading the fencer's matches together for comparison}
	\label{fig:comparison}
\end{figure*}

\subsection{Comparison of Three Bouts}

On the basis of the first case, we compare the tactical flow graphs of the two semifinal and final matches of the Men's Sabre Individual Golden Match of the 2017 Fencing World Championship. For ease in comparison, we switch the position of Gu and Iburagimov for the semi-finals, such that the two fencers (Szatmari and Gu) are in the same position on the graph in their respective two games. 

The thickest flow always immediately catches the attention of viewers. The \textit{S-FF} flow is the thickest in all three graphs, which is consistent with the dominant position of attacks in sabre match. In addition, the \textit{S-BB} flow in the finals is significantly thicker than the corresponding flows in the two semi-finals, which indicate that the fencers played more conservatively and chose to retreat more frequently in the finals.

By comparing the flow at the end of the bouts, we can see that the number of forward points lost by the winners of all matches is relatively small, for example, FB-2 in A and B and BF-1 in C are relatively thin. 
In addition, the FF-1 flow in C is also relatively thin, which shows that Gu has stronger ability to score when both fencers advancing.

In addition, the flow of FB-BF and BF-FB in relatively thin in all the three views, which shows that the main winning method in the sabre competition is to attack directly or attack folled a fake retreat, and the situation of attack-defense conversion is relatively small and it rarely shows obvious advantages. This is obvious especially in high-level competitions,  basic skills of fencers are generally equal. 

The FencingVis system can easily compare different games given the fencing scenarios mentioned above, and such ease in comparison is not available in previous work
\cite{wu2018ittvis,polk2014tennivis}.

\subsection{Comparison of Different Matches of The Same Fencer}
Our system also supports the analysis of the characteristics of a fencer by loading the fencer's matches together for comparison, as shown in \autoref{fig:comparison}.

Gu's latest three bouts were shown in \autoref{fig:comparison}A.
For the sake of comparison, we have changed Gu's position in the three match all to the left. 
The blue flow represents his only win of the three games.
We can find some interesting patterns in the view.
First of all, in the match Gu wins, it is not occurred that he chooses forward tactic with his opponent retreat and scored, though which occurred several times in the other two matches (shown in \autoref{fig:comparison}A-2).
According to this phenomenon, we can judge the attacking been riposted is Gu's weakness, and if the opponent will have a greater chance of winning if he caught this.
In addition, we also found Gu get significant more with both sides choosing forward tactics than the other two matches (\autoref{fig:comparison}A-3).
This is consistent with our previous analysis, Gu's offensive ability is powerful, and his opponent will suffer if he chooses to play a hard ball. But if the opponent chooses to retreat to riposte, his probability of scoring will be higher.
In addition, we found that there was no \textit{BF-FB} and \textit{BB-FB} transition in Gu's winning game (\autoref{fig:comparison}A-1).
The two flows both means the fencer gaining priority.
They were supposed to be the embodiment of the superiority tactics, but has played the opposite effect for Gu.
We have concluded that Gu's ability to turn forward after retreating is not good, and often easy to fall into the trap set by the opponent. 
We can see this in \autoref{fig:comparison}A-2, he is supposed to lose point after gain priority in this situation.
Based on the above analysis, we can get a very clear conclusion.
Gu's offensive ability is strong, retreat ability is weak, and if we encountered this opponent, must not play hardball, pull back is a better way than direct attack.

Then let's take a look at Szatmari's two matches, both of which he wins, as shown in \autoref{fig:comparison}B.
One of the most obvious characteristics is the \textit{BF-FB} flow's absence (\autoref{fig:comparison}B-1).
This flow reflects the fencer's succefully handling the opponent's attack and gaining priority.
As we can see, this is Szatmari's short board.
Once the opponent gets the priority, he is hard to take it back.
But to make up, he always can avoid falling into long-range offensive or defensive at the beginning of each phrase by using appropriate tactics.
The proper use of tactics made up for his lack of ability in this area, making it difficult to find a very effective way to beat him, so he won the championship at last.

\section{Discussion}
In the design and development of FencingVis, we refer to the developmental process model proposed by Sedlmair et al.\cite{sedlmair2012design}. However, substantial problems need to be clarified. First, fencing data comprise two related time-series data. We initially planned to adopt the time-series data analytical method, an approach widely accepted for fencing data. However, upon further understanding of the problem, we found that the time-series data of fencing involve a hierarchical structure. The motion-series data at different stages of a phrase also involve different technical and tactical information. For example, two fencers always choose to move forward at the beginning of the phrase, and only one-step or two-steps forward is observed in this stage according to our statistics. The one-step and two-step movement in the initial stage of a phrase is closely related to the technical characteristics of fencing and the tactics selected by fencers, and both greatly influence succeeding competitions. By contrast, after entering the long-distance attack and defense, the number of forward and backward steps is often unimportant, but the depth of the lunge will be more important than the timing. As such, we conduct a two-level analysis. 

We first represent the time-series data of a phrase as a sequence of tactical-combination behavior corresponding to higher-level abstraction, which reflects the tactical application of both fencers in a specific phrase. Then, for each node of the sequence, we analyze the technical ability, such as reaction time and attack position, of both fencers. Previous studies mainly focused on analyzing the technical capabilities of a sequence in its entirety. However, according to our research, additional detailed patterns and features can be found by analyzing the data using a multi-layer framework. On the basis of this hierarchical structure, the data presentation and its interactive features are designed as the two abstract levels of tactical information and technical information, respectively.

However, our focus is still on the tactical level. Substantial work has been conducted to analyze the technical characteristics of fencing. By proposing FencingVis, we hope to offer users with a new perspective of analysis based on the abovementioned tactical framework so that the technical characteristics of fencers can be easily understood.

In the process of system design and development, we cooperate deeply with three experts in fencing field. 
They were all elite fencers before, and teach in universities and professional teams now.
Many of their professional suggestions give us a lot of inspiration.
They also suggest that our system, apart from analyzing professional competitions, can also be used to teach fencing beginners or demonstrate tactics to fencing enthusiasts.

\section{Conclusion}
We design and implement a visual analysis system called FencingVis for visualization and visual analysis of fencing data. 
We use multiple views to present the data from different perspectives and provide exploratory analysis methods to domain experts through a series of interactive filters and view coordination.
By using three case studies as basis, we prove that FencingVis can help domain experts find the patterns that were previously difficult to discover. The experts have also shared substantial positive feedback for the system.

At present, FencingVis is mainly aimed toward the individual matches of sabre. Given that the rules for epee and foil slightly differ, the system needs to be further improved to accommodate the other two principles. Similarly, we also plan to improve our system to support the team events.
Team events involve more fencers, the ordering of the fencers adds a new dimension to the analysis, which is a problem to be solved in our future work.

\acknowledgments{
	This research is partially supported by National Natural Science Foundation of China (Grant Nos. 61572274, 61672307, 61272225) and the National Key R\&D Program of China (Grant No. 2017YFB1304301).
}

\bibliographystyle{abbrv-doi}

\bibliography{FencingVis}
\end{document}